\documentclass[12pt]{article}
\usepackage{cite,psfig}

\newcommand{\newc}{\newcommand}
\newc{\gsim}{\lower.7ex\hbox{$\;\stackrel{\textstyle>}{\sim}\;$}}
\newc{\lsim}{\lower.7ex\hbox{$\;\stackrel{\textstyle<}{\sim}\;$}}
\newc{\gev}{\,{\rm GeV}}
\newc{\mev}{\,{\rm MeV}}
\newc{\ev}{\,{\rm eV}}
\newc{\kev}{\,{\rm keV}}
\newc{\tev}{\,{\rm TeV}}
\def\ln{\mathop{\rm ln}}

\newc{\mz}{M_Z}
\newc{\mpl}{M_*}
\newc{\mw}{m_{\rm weak}}
\def\beq{\begin{equation}}
\def\eeq{\end{equation}}
\def\bea{\begin{eqnarray}}
\def\eea{\end{eqnarray}}
\newc{\ie}{{\it i.e.}}          \newc{\etal}{{\it et al.}}
\newc{\eg}{{\it e.g.}}          \newc{\etc}{{\it etc.}}
\newc{\cf}{{\it c.f.}}
\def\bar#1{\overline{#1}}

\def\inv{^{\raise.15ex\hbox{${\scriptscriptstyle -}$}\kern-.05em 1}}
\def\lbar{{\lower.35ex\hbox{$\mathchar'26$}\mkern-10mu\lambda}} 

\def\to{\rightarrow}

\let\<=\langle
\let\>=\rangle

\let\+=\uparrow
\renewcommand{\epsilon}{\varepsilon}
\renewcommand{\phi}{\varphi}
\def\mfo{M_4}
\def\mfi{M_5}
\def\mi{M_c}
\def\rh{r_h}
\def\lc{\ell}
\def\rd{\rho_d}

\begin{document}
\thispagestyle{empty}
\vspace*{.5cm}
\noindent
\hspace*{\fill}{\large CERN-TH/2001-078}
\vspace*{2.0cm}

\begin{center}
{\Large\bf Randall-Sundrum II Cosmology,\\[.2cm] AdS/CFT, and the Bulk Black 
Hole}
\\[2.5cm]
{\large Arthur Hebecker and John March-Russell
}\\[.5cm]
{\it Theory Division, CERN, CH-1211 Geneva 23, Switzerland}
\\[.2cm]
(March 20, 2001)
\\[1.1cm]

{\bf Abstract}\end{center}
\noindent
We analyse the cosmology of a brane world model where a single brane 
carrying the standard model fields forms the boundary of a 5-dimensional 
AdS bulk (the Randall-Sundrum II scenario).  We focus on the thermal 
radiation of bulk gravitons, the formation of the bulk black hole, and the 
holographic AdS/CFT definition of the RSII theory.  Our detailed calculation 
of bulk radiation reduces previous estimates to a phenomenologically 
acceptable, although potentially visible level.  In late cosmology, in which 
the Friedmann equation depends linearly on the energy density $\rho$, only 
about 0.5\% of energy density is lost to the black hole or, equivalently, to 
the `dark radiation' ($\Omega_{d,N} \simeq 0.005$ at nucleosynthesis).  The 
preceding, unconventional $\rho^2$ period can produce up to 5\% dark 
radiation ($\Omega_{d,N} \lsim 0.05$).  The AdS/CFT correspondence provides
an equivalent description of late RSII cosmology.  We show how the AdS/CFT
formulation can reproduce the $\rho^2$ correction to the standard treatment
at low matter density.  However, the 4-dimensional effective theory of CFT
+ gravity breaks down due to higher curvature terms for energy 
densities where $\rho^2$ behaviour in the Friedmann equation is usually 
predicted.  We emphasize that, in going beyond this energy density, the
microscopic formulation of the theory becomes essential.  For example,
the pure AdS$_5$ and string-motivated AdS$_5\times$S$^5$ definitions
differ in their cosmological implications. 
\newpage

\section{Introduction}

The brane world scenarios of Randall and Sundrum \cite{rs1,rs2} have opened 
new perspectives in physics beyond the standard model. The one-brane 
scenario of~\cite{rs2} (RSII) appears to be particularly interesting from 
the point of view of gravitation  and cosmology. 

In RSII, the standard model fields are confined to the boundary of an 
infinitely extended AdS$_5$ bulk. Four-dimensional gravity is recovered at 
length scales above the AdS length $\lc$. While related models with warped 
and non-compact extra dimensions have been considered before~\cite{old}, 
an extensive discussion has been triggered by the specific model 
of~\cite{rs2}. An important reason for this is the proximity of this 
construction to recent more formal developments, such as D-branes and the 
AdS/CFT correspondence. On the phenomenological side, the RSII model 
demonstrates the extent to which gravity can be changed without violating 
experimental constraints. This is manifest in the existence of infinitely 
many, weakly coupled light modes and in the possibility of an 
unconventional cosmological evolution. 

Although there have been many previous treatments of 
`brane cosmology', the majority of these focus on either the 
situation for flat extra dimensions~\cite{add}, which is highly 
constrained because of either the problem of bulk gravitons or the radion 
problem~\cite{flatcosmo}, or on the situation in RSI with the additional 
`TeV-brane' and radion problems~\cite{rs1cosmo}.

In the present paper, we develop the physical picture of the early 
universe in RSII, including in particular the phenomenologically
relevant dynamics of bulk gravitons, the physics of the bulk black hole,
the use of the AdS/CFT correspondence to make manifest some
properties of the theory in the IR, and the varying early cosmology
resulting from the possibility of quite different definitions of the theory
in the UV.  

The investigation of RSII cosmology and phenomenology was pioneered 
in~\cite{bin,kra,sms,gub} and~\cite{rs2p,gkr,ced,gk} respectively. The 
cosmological evolution is most conveniently discussed in terms of a 
moving brane with static AdS-Schwarzschild metric in the bulk. This is 
equivalent to the description with an expanding bulk~\cite{msm}. The 
two main unconventional cosmological features are the dark radiation 
(characterized by the size of the black hole in the bulk) and the 
$\rho^2$ term in the Friedmann equation. Both effects are governed by the 
`AdS-cosmology scale' $\mi=\sqrt{\mfo/\lc}$, where $\mfo$ is the 
4-dimensional reduced Planck mass. In the most radical scenario 
with $\lc\sim 1$ mm, one has a 5-dimensional reduced Planck mass 
$\mfi=\sqrt[{\scriptstyle 3}]{\mfo^2/\lc}\simeq 1.1\times 10^5$ TeV and 
$\mi\simeq 0.7$ TeV. 

So far, much attention has been devoted to the geometrical brane dynamics 
and its possible extensions~\cite{geo} and to the evolution of 
inhomogeneities in the early universe~\cite{inho} (see~\cite{rev} for 
recent reviews). However, the discussion of radiation of gravitons
by the brane has been less prominent. An estimate of this effect has been 
given in~\cite{gub}. Other, model dependent, sources for dark radiation 
have been discussed in~\cite{ckn}.

We argue that graviton radiation by the hot brane will unavoidably produce 
a black hole in the bulk. If one assumes that the black hole was initially
very small (e.g., because of inflation on the brane), then the size of the 
black hole is a calculable function of the reheating temperature. We present 
a detailed calculation of the rate at which energy is lost to the bulk due
to graviton radiation. Our result allows for an extended $\rho^2$ period in 
the early universe. More specifically, we find that the $\rho$ period 
produces dark 
radiation corresponding to $\Omega_{d,N}\simeq 0.005$ at nucleosynthesis, 
while a maximally extended $\rho^2$ period can result in $\Omega_{d,N}\sim 
0.05$. These numbers are small since, parametrically, they are inversely
proportional to the number of brane degrees of freedom (this number, which 
is ${\cal O}(100)$ in the standard model, enhances the expansion rate, thus 
leaving less time for graviton production). The contribution of the 
$\rho^2$ period is enhanced by a factor $\sim\ln(T_{\mbox{\scriptsize max}}
/\mi)$ with the respect to the $\rho$ period, where $T_{\mbox{\scriptsize 
max}}$ is the maximal brane temperature. It is interesting to note that 
gravitons emitted sufficiently early remain bound to the brane and fall
into the horizon only at the end of the $\rho^2$ period. Our numerical 
results for $\Omega_{d,N}$ can be lowered or enhanced by having additional 
degrees of freedom on the brane or in the bulk respectively. From a
phenomenological point of view, it is amusing to note that the size of 
$\Omega_{d}$ is potentially observable. 

Gubser~\cite{gub} has initiated an AdS/CFT description of RSII cosmology 
(see also~\cite{rs2cft}) in 
which the CFT corresponds to the bulk degrees of freedom. The CFT 
temperature, which is the bulk temperature near the brane, is responsible 
for the dark radiation. It is known that the leading corrections to 4d 
gravity arising from the bulk Kaluza-Klein modes are reproduced by the 
CFT-gravity coupling. We point out that the direct coupling of brane fields 
to the CFT can be responsible for the known $\rho^2$ corrections in the 
Friedmann equation. In addition, at the same densities when these 
corrections become relevant, higher curvature terms induced by the CFT start 
to affect the 4d gravity theory. These effects, naively suppressed by 
$\mfo$, come in so early because of the large number of degrees of freedom 
of the CFT. Furthermore, we emphasize that the `UV region' $r<\lc$ of RSII 
is very different in the string motivated case of AdS$_5\times$S$^5$.

Our paper is organized as follows. After discussing the basic cosmological 
setting and the origin of the bulk black hole in Sect.~2, we derive the 
energy loss through graviton radiation in Sect.~3. Section~4 deals with 
the AdS/CFT perspective and, in particular, with the UV definition of 
RSII. We summarize and outline interesting open questions in Sect.~5 
and present some formulae related to the graviton production rate in the 
Appendix.

\section{Basics of late cosmological evolution}
We start from the 5-dimensional action~\cite{rs2} 
\beq
S=\int d^5x \sqrt{-g_5} \left(\frac{1}{2}\mfi^3{\cal R}-\Lambda_5\right)
\,\,+\,\,\int_{brane}d^4x\sqrt{-g_4}({\cal L}_{SM}-\Lambda_4)\,,\label{act}
\eeq
which includes the contribution from a 4-dimensional brane with induced
metric $g_4$. We suppose that the standard model fields characterized by 
the Lagrangian ${\cal L}_{SM}$ are localized on this brane. The situation is 
further 
simplified by identifying the regions on both sides of the brane, i.e., 
imposing $Z_2$ orbifold boundary conditions on the 5-dimensional space-time.
Our treatment of the dynamics of this system will follow closely the 
particularly clear and compact discussion of Kraus~\cite{kra}. 

The motion of the brane follows from the 5-dimensional Einstein equations. 
It can be characterized most conveniently in terms of the extrinsic 
curvature $K_{\mu\nu}$ of the brane (with brane indices $\mu,\nu=0...3$). 
$K_{\mu\nu}$, also known as the second fundamental form, can be defined as 
the projection of 
\beq
K_{MN}=\nabla_M n_N\,,
\eeq
(with bulk indices $M,N=0...4$) onto the brane, where $n_N$ is the unit 
normal. Then the brane motion is determined by the Israel junction 
condition~\cite{is} (see also~\cite{bgg})
\beq
K^+_{\mu\nu}-K^-_{\mu\nu}=2K_{\mu\nu}=-\frac{1}{\mfi^3}\left(T_{\mu\nu}-
\frac{1}{3}T_\rho^\rho\,g_{4,\mu\nu}\right)\,,\label{is}
\eeq
where the index `+' refers to the side of the brane to which $n^N$ 
points and the index `$-$' refers to the opposite side of the brane. 
The energy-momentum tensor $T_{\mu\nu}$ follows from the brane action in 
Eq.~(\ref{act}). 

For negative $\Lambda_5$, the AdS metric 
\beq 
ds^2=-\frac{r^2}{\lc^2}dt^2+r^2(d\vec{x})^2+\frac{\lc^2}{r^2}dr^2\,, 
\label{ads}
\eeq
with $\lc^2=6\mfi^3/|\Lambda_5|$, represents a vacuum solution of the 
5-dimensional Einstein equations. If, in addition, the 4-dimensional 
cosmological constant, i.e., the brane tension, satisfies the relation 
$\Lambda_4=6\mfi^3/\lc$, then the Israel junction conditions allow for
an empty brane to be at rest in the above coordinates, e.g., at 
the position $r=R$. 

In this situation, the brane observer sees 4-dimensional gravity at 
distance scales greater than $\lc$. The corresponding reduced Planck mass, 
$\mfo=1/\sqrt{8\pi G_4}$, is fixed by the relation
\beq
\mfo^2=\mfi^3\lc\,,
\eeq
where $\mfi$ is the analogously defined 5-dimensional reduced Planck mass. 
Thus, in our following analysis, it will be convenient to consider $\mfo$ 
and $\lc$ as our basic physical parameters and $\mfi$, $\Lambda_4$ and 
$\Lambda_5$ as derived quantities. 

Starting from Gauss's Theorema Egregium (see, e.g.,~\cite{nic}), which 
expresses the bulk curvature near the brane in terms of the intrinsic and 
extrinsic brane curvatures, and replacing the extrinsic brane curvature 
with the brane energy-momentum tensor according to Eq.~(\ref{is}), one 
arrives at~\cite{sms} (see also~\cite{ha})
\beq
\mfo^2{\cal R}^4_{\mu\nu}=\tau_{\mu\nu}-\frac{1}{2}\tau g_{4,\mu\nu}-\frac{1}
{4\mi^4}\tau_\mu^\rho\left(\tau_{\rho\nu}-\frac{1}{3}\tau g_{4,\rho\nu}
\right)+\mfo^2\delta{\cal R}^5_{\mu\rho\nu\sigma}g_4^{\rho\sigma}\,.
\label{ei}
\eeq
Here $\tau_{\mu\nu}=T_{\mu\nu}+\Lambda_4g_{4,\mu\nu}$ is the matter 
contribution to the brane energy-momentum tensor, ${\cal R}^4_{\mu\nu}$ is 
the brane Ricci tensor and $\delta{\cal R}^5_{\mu\rho\nu\sigma}$ is the 
brane projection of the deviation of the bulk curvature tensor from its 
pure-AdS value. As will be discussed in detail below, the `AdS-cosmology 
scale'
\beq
\mi=\sqrt{\mfo/\lc}
\eeq
plays a prominent role in the cosmology of the Randall-Sundrum model. It 
is intermediate with respect to the AdS-scale $1/\lc$ and the `strong 
gravity scale' $\mfi=\sqrt[{\scriptstyle 3}]{\mfo^2/\lc}$ since, in the 
physically interesting limit $1/\lc\ll\mfo$, one has $1/\lc\ll\mi\ll\mfi$. 
Equation~(\ref{ei}) reduces to the Einstein equation for the induced brane 
metric in the limit where the bulk is pure AdS and energy densities on the 
brane are much smaller than $\mi^4$. 

To discuss cosmology, we have to introduce hot matter on the brane. In this 
situation, the bulk solution, Eq.~(\ref{ads}), is not general enough. The 
reason for this is the unavoidable graviton radiation off the brane. No 
matter how small the amount of energy lost to the bulk might be, it will 
fall deep into the AdS bulk and produce, at least if back-reaction on the 
metric is neglected, an arbitrarily high local energy density. The only 
plausible physical resolution of this problem that we are aware of is 
offered by the creation of a black hole in the bulk.

To see this more explicitly, consider the trajectory of a graviton
radiated 
at time $t=0$ by a static brane at position $R$. At large $t$, this 
trajectory is given by 
\beq
r=\frac{R}{1+(Rt/\lc^2)\sin\phi}\simeq\frac{\lc^2}{t\sin\phi}\,,
\eeq
where $\phi$ is the angle between the initial graviton momentum and the 
brane. (The dependence on this angle is obtained most easily by
considering 
brane-perpendicular radiation first and then boosting the configuration 
along the brane.) Assume next that a 3-dimensional energy density 
$\Delta\rho$ is lost to gravitons with angles between $\phi$ and $\phi+
\Delta\phi$. At large $t$, these gravitons will be found in a slice of 
the AdS bulk of invariant thickness
\beq
\frac{\lc}{r}\,\Delta r\simeq \frac{\lc}{r}\cdot\frac{\lc^2\,\cos\phi}{t\,
\sin^2\phi}\cdot\Delta\phi\simeq\frac{\cos\phi}{\sin\phi}\cdot\lc\,\Delta
\phi\,.
\eeq
Note that the separation of gravitons in $r$ direction due to different 
times of their emission from the brane is small compared to the separation 
due to different angles of emission. In the rest frame of the matter on
the 
brane, the brane-parallel components of the graviton momenta are 
isotropically distributed. Therefore the radiation in the above slice 
$\Delta r$ has its own `rest frame'. In this frame, the momenta in $r$ 
direction vanish and the momenta in brane direction are blue-shifted by a 
factor $R/r$ with respect to their original values. Furthermore, the
density 
in brane direction is increased by a factor $(R/r)^3$ due to AdS geometry. 
Thus, the 4-dimensional energy density of the radiation at position $r$, 
measured in its own rest frame, is given by 
\beq
\rho_r\simeq\frac{\Delta\rho}{(\lc/r)\,\Delta r}\,\left(\frac{R}{r}
\right)^4\,\cos\phi\simeq\frac{\Delta\rho}{\lc\,\Delta\phi}\,\left(\frac{R}
{r}\right)^4\,\sin\phi\,.
\eeq
This local energy density diverges as $r\to 0$ and we are forced to 
conclude that eventually a horizon will form, hiding any possible 
super-Planck-scale effects from our view. 
Thus, the relevant bulk metric is given by the AdS-Schwarzschild 
solution
\beq
ds^2=-f(r)dt^2+r^2(d\vec{x})^2+f(r)^{-1}dr^2\,
\label{ds}
\eeq
with
\beq
f(r)=\frac{r^2}{\lc^2}\left(1-\frac{\rh^4}{r^4}\right)\,,
\eeq
where $\rh$ is the position of the black hole horizon. 

Now the Israel junction conditions or, equivalently, Eq.~(\ref{ei}) lead to 
the following equation of motion for the brane:
\beq
3\mfo^2\left(\frac{\dot{R}}{R}\right)^2=\rho\,\left(1+\frac{\rho}{12\mi^4}
\right)+3\mi^4\left(\frac{r_h}{R}\right)^4\,,\label{hu}
\eeq
where $\dot{R}=dR/d\tau$ and $\tau$ is the proper time of the brane 
observer. Since the brane metric is given by 
\beq
ds^2_b=-d\tau^2+R^2(d\vec{x})^2\,,
\label{dsb}
\eeq
the small-$\rho$ and small-$\rh$ limit of Eq.~(\ref{hu}) reproduces the 
familiar 4-dimensional Friedmann equation. Deviations are characterized 
by $\mi$. On the one hand, $\mi$ determines the scale at which the $\rho^2$ 
term in Eq.~(\ref{hu}) becomes important. On the other hand, it sets the 
scale of the last term on the rhs of Eq.~(\ref{hu}). This term, which 
contributes to the expansion rate like an energy density $\rd$ of `dark 
radiation', is $\sim\mi^4$ if the brane is near the black hole horizon, 
$R\sim\rh$. Furthermore, as will be discussed in detail in the next section, 
the radiation of bulk gravitons competes with the Hubble expansion rate 
$H=\dot{R}/R$ at brane temperatures $T\gsim\mi$. 

Let us add a comment concerning the position of the horizon (or, 
equivalently, the size of the black hole) characterized by $r_h$. As can
be 
seen from Eqs.~(\ref{ds})--(\ref{dsb}), the absolute value of this 
quantity is not physical. In fact, nothing changes if one rescales the 
bulk coordinates and the positions of brane and horizon according to
\beq
x^\mu\to \alpha\,x^\mu\quad,\qquad r\to r/\alpha\qquad\mbox{and}\qquad 
R\to R/\alpha\quad,\qquad r_h\to r_h/\alpha\,,
\eeq
with some real number $\alpha>0$. The important physical parameter is the 
ratio $r_h/R$ (see Eq.~(\ref{hu})). Note furthermore that this degeneracy
is 
lifted if one considers closed or open geometries, where $f(r)=k+r^2/\lc^2
(1-r_h^2/r^2)$ with $k=\pm 1$. However, since we are only interested in
the early universe, we will primarily consider the flat case with $k=0$.

\section{Dark radiation and graviton emission into the bulk}

Let us start with the cross section for the production of bulk gravitons 
by matter on the brane. If the cms-energy of the collision is large 
compared to particle masses, $\sqrt{s}\gg m$, and to the typical scale of 
the AdS-space curvature, $\sqrt{s}\gg 1/\lc$, the only relevant scales in 
the process are $\sqrt{s}$ and the 5-dimensional Planck mass. Since the 
cross section has to be proportional to the 5-dimensional gravitational 
coupling, we expect $\sigma(s)\sim\sqrt{s}/\mfi^3$ on dimensional
grounds. \footnote{
This result differs from what is implied by Eq.~(33) of \cite{gub} (see also
our Eq.~(\ref{de})). The discrepancy arises because the estimate of 
Ref.~\cite{gub} is based on the leading corrections to 4-dimensional 
gravity. By contrast, we find that the relevant regime is $\sqrt{s}\gg 
1/\lc$, i.e., truly 5-dimensional gravity. As a result, we get weaker 
bounds and high allowed brane temperatures.}

The exact results, which are derived in the Appendix utilizing previous work 
of~\cite{ced,grw}, can be summarized by writing
\beq
\sigma_i(s)=c_i\,\frac{\sqrt{s}}{\mfi^3}\,.\label{cs}
\eeq
The appropriate numerical constants for the cases of scalars, vector 
particles and fermions (with initial-state spin averaging included) are 
given by
\beq
c_s=\frac{1}{12}\qquad,\qquad c_v=\frac{1}{4}\qquad,\qquad 
c_f=\frac{1}{16}\,.\label{cfa}
\eeq

In a hot plasma, the reaction rate per 3-volume is obtained by thermally
averaging the cross section $\sigma$ for the process under consideration 
and multiplying it with the squared number density $n$ of the initial state 
particles (see, e.g., Ref.~\cite{swo}). Analogously, the total rate of 
energy loss due to bulk graviton radiation is obtained by thermally 
averaging the product of cross section and lost energy (to avoid double 
counting the number of interactions between identical particles, a factor
$1/2$ has to be included):
\beq
\Delta\dot{\rho}=-\frac{1}{2}\langle\sigma v_{\mbox{\scriptsize rel}}\,
\cdot\,E\rangle\,n^2=-\frac{1}{2}\int d^3p_1\,d^3p_2\,f(E_1)f(E_2)\,\,\sigma 
v_{\mbox{\scriptsize rel}}\,\cdot\,(E_1+E_2)\,.\label{ee}
\eeq
Here $f(E)=\kappa(2\pi)^{-3}(\exp(E/T)\pm 1)^{-1}$ is the distribution 
function for bosons (minus sign) or fermions (plus sign) with $\kappa$ 
spin degrees of freedom and $v_{\mbox{\scriptsize rel}}$ is the relative 
velocity of the colliding particles. Using the cross section of 
Eq.~(\ref{cs}) and setting particle masses to zero, one finds
\beq
\Delta\dot{\rho}=-\frac{2T^8c_i}{5\pi^4\mfi^3}\kappa_i^2\,\textstyle\Gamma(
\frac{7}{2})\zeta(\frac{7}{2})\Gamma(\frac{9}{2})\zeta(\frac{9}{2})\,,
\label{de}
\eeq
in the bosonic case. An additional factor $a_f=(1-2^{-5/2})(1-2^{-7/2})
\simeq 0.750$ arises in the fermionic case. 

Summing the different particle species and normalizing to the total energy 
density of a relativistic gas one obtains
\beq
\frac{\Delta\dot{\rho}}{\rho}=-C\frac{T^4}{\mfi^3}\,,\label{dec}
\eeq
where
\beq
C=0.574\,\,\cdot\,\frac{g_sc_s+2g_vc_v+2g_fa_fc_f}{g_s+g_v+(7/8)g_f}\,,
\label{cdef}
\eeq
for $g_s$ scalar, $g_v$ vector, and $g_f$ fermionic degrees of freedom. 
The factors 2 in front of $g_v$ and $g_f$ are due to the 2 spin degrees
of freedom of massless vector particles and fermions. In the standard 
model, $g_s=4$, $g_v=24$ and $g_f=90$, leading to $C=0.112$. 

At sufficiently late times, the expansion of the universe is governed by
4-dimensional gravity. To be more specific, this period is characterized 
by $1/H\gg\lc$ or $\rho\ll\mi$, which means that the term linear in $\rho$ 
dominates the rhs of Eq.~(\ref{hu}) (`$\rho$ period'). In this 
situation, effective 4-dimensional energy-momentum conservation ensures 
that the loss of energy-density on the brane is equal to the gain of dark 
energy-density $\rd$,
\beq
\Delta\dot{\rho}+\Delta\dot{\rho}_d=0\,,
\eeq
where both $\rho$ and $\rd$ scale like radiation: $\dot{\rho}=-4H\rho+ 
\Delta\dot{\rho}$ and $\dot{\rho}_d=-4H\rd+\Delta\dot{\rho}_d$. Thus, even 
if $\rd=0$ at some initial time $\tau_1$, it will develop a non-zero 
late-time value characterized by $\Omega_d=\rd/(\rho+\rd)$. If $\rd\ll\rho$, 
we have
\beq
\Omega_d=\int_{\tau_1}^{\infty}d\tau\,\left(-\frac{\Delta\dot{\rho}}{\rho}
\right)\,,\label{omi}
\eeq
where the integrand is given by Eq.~(\ref{dec}). Employing the relation 
$\rho=g_*(\pi^2/30)T^4$ (where $g_*$ is the effective number of degrees of 
freedom, $g_*=g_s+g_v+(7/8)g_f$ ), the temperature factor $T^4$ can be 
expressed through $\rho$. Now the integral in Eq.~(\ref{omi}) is easily 
performed, giving the result 
\beq
\Omega_d=\frac{15C}{g_*\pi^2}\,\sqrt{\frac{3\rho_1}{\mi^4}}\,,\label{om}
\eeq
where $\rho_1$ is the radiation density at $\tau=\tau_1$. The largest value 
of $\rho_1$ compatible with linear $\rho$ behaviour is $\rho_1\simeq 
12\mi^4$ (cf.~Eq.~(\ref{hu})), leading to
\beq
\Omega_d=\frac{90C}{g_*\pi^2}\,.\label{ni}
\eeq
Assuming that the evolution of the universe after the decoupling of bulk 
gravitons (which is complete soon after the beginning of the $\rho$ period) 
respects entropy conservation, one finds that at the time of 
nucleosynthesis
\beq
\Omega_{d,N}=\left(\frac{g_{*,\,N}}{g_*}\right)^{1/3}\,\Omega_d\,.
\eeq
Here $g_{*,\,N}\simeq 10.75$ is the number of light degrees of freedom at
nucleosynthesis. Thus, one arrives at the result that bulk graviton 
production during the complete $\rho$ period only gives rise to 
$\Omega_d=0.0044$, i.e., somewhat less than 0.5\% of energy density in dark 
radiation. The smallness of this contribution is a direct result of the 
large number of light fields on the brane, which leads to a fast expansion 
and a correspondingly fast cooling of the universe. 

The above result is encouraging for two reasons. First it implies that
late cosmological evolution in the RSII model is quite safe (although
not without a potential signature in future accurate CMBR measurements)
even assuming the most extreme values for $\lc$ and $\mfi$. Second it 
allows for a cosmological $\rho^2$ period with hot matter on the brane. 
However, also during the 
$\rho^2$ period matter is lost to dark radiation and we have to calculate 
the resulting contribution to $\Omega_d$. Since we do not have effective 
4-dimensional gravity during the $\rho^2$ period, the energy lost by the 
brane is not necessarily the same as the energy found at late times in 
dark radiation, $\Omega_{lost}\neq\Omega_d$. Equation~(\ref{omi}) is still
valid, but with the lhs replaced by $\Omega_{lost}$ and with $\tau_1$ 
as the upper limit of integration,
\beq
\Omega_{lost}=\int_{\tau_0}^{\tau_1}d\tau\,\left(-\frac{\Delta\dot{\rho}}
{\rho}\right)\,.\label{om2}
\eeq
Here $\tau_0$ is the earliest time at which our analysis is valid, e.g., the 
time of reheating. 

The Friedmann equation in the $\rho^2$ period,
\beq
6\mfi^3H=\rho\,,
\eeq
which follows from Eq.~(\ref{hu}) in the limit $\rho\gg\mi^4$, implies that
$\rho\sim 1/\tau$ and $H=(1/4\tau)$. In this situation, Eqs.~(\ref{dec}) and 
(\ref{om2}) give
\beq
\Omega_{lost}=\frac{45C}{g_*\pi^2}\ln(\rho_0/\rho_1)\simeq
\frac{90C}{g_*\pi^2}\,\cdot\,\frac{1}{3}\ln\left(M_4\lc/12^{\,3/2}\right)\,.
\eeq
The last approximate equality follows by assuming $\rho_0\simeq\mfi^4$, the 
highest density compatible with weakly interacting gravity. This result 
shows an enhancement by a potentially large factor $(1/3)\ln(M_4\lc)$ 
compared to Eq.~(\ref{ni}). 

We still have to adress the question of how $\Omega_{lost}$ at the $\rho^2$ 
period translates into $\Omega_d$ at late times. This is most easily done 
from the brane perspective. During the $\rho^2$ period, the brane moves 
with almost speed-of-light through the AdS bulk (a natural bulk rest frame 
can be defined by the black-hole horizon or, equivalently, by the late-time
limit, where the brane is almost at rest). The (absolute) acceleration of 
the brane can be calculated using the standard formula 
\beq
a^M=u^N\nabla_N u^M\,,
\eeq
where $u^M=dx^M/d\tau$ is the velocity of a particle at rest on the 
brane. Multiplying this acceleration with the brane normal vector $n^M$ 
(which is, in fact, the direction of the acceleration) and making use of 
the orthogonality relation $n\cdot u=n_M u^M=0$, one derives
\beq
a=n\cdot a= -u^M u^N\nabla_M n_N\,.
\eeq
This is precisely the $\tau\tau$ component of the extrinsic curvature 
$K_{\mu\nu}$. Using the Israel junction condition, Eq.~(\ref{is}), and 
defining $n^M$ to point into the direction of increasing $r$, the 
acceleration is found to be 
\beq
a=-\frac{1}{2\mfi^3}\,(\Lambda_4+\rho)\,.
\eeq
Notice first that this provides a nice physical interpretation of the 
Israel junction condition: the brane motion is determined by the brane 
acceleration, which, in turn, is determined by the energy density on the 
brane. Furthermore, deep in the $\rho^2$ period, the brane can be thought 
of as accelerating into the direction of decreasing $r$ with $a\sim
\rho/\mfi^3$. This implies that gravitons radiated by the brane are 
accelerated towards the brane with the same acceleration $a$. Since, in a 
thermalized situation, gravitons are emitted with smooth angular 
distribution, most of them will fall back onto the brane after having 
reached a maximal distance 
\beq
d\sim \frac{1}{a}\sim\frac{\mfi^3}{\rho}\,. \label{rcv}
\eeq
Clearly, this discussion becomes invalid if $d\gsim\lc$, since then the 
AdS curvature takes over and determines the future path of the graviton. 
In this situation, the graviton will not return to the brane but fall into 
the black hole horizon. 

The critical value of $\rho$ is determined Eq.~(\ref{rcv}) with $a\sim 
1/\lc$, 
\beq
\rho\sim \frac{\mfi^3}{\lc}\sim\mi^4\,.
\eeq
This means that virtually no gravitons leave the vicinity of the brane 
during the $\rho^2$ period. Instead, the gravitons radiated by the brane 
remain gravitationally bound to the brane and fall into the black hole 
only at the end of the $\rho^2$ period, when the motion of the brane 
becomes non-relativistic. 

Between its initial emission and its final fall into the horizon, each
graviton can bounce off the brane many times. During this process, the 
graviton momentum parallel to the brane remains unchanged except for the 
trivial redshift factor associated with the AdS geometry. However, the 
momentum transverse to the brane decreases dramatically. This is seen most
easily by observing that, in the black hole frame, the graviton is 
reflected many times by the retreating brane, losing part of its momentum 
in the $r$ direction with each reflection. 

Thus, we expect that the $\rho^2$ period contributes an amount $\Omega_d=
\alpha\Omega_{lost}$ (with $\alpha<1$) to the late-time dark energy 
fraction. A lower bound on the constant $\alpha$ can be derived by 
assuming that only the energy going into the brane-parallel motion of the 
radiated gravitons will survive the multiple reflection phase and make it 
into the black hole horizon. This corresponds to replacing the factor 
$(E_1+E_2)$ in Eq.~(\ref{ee}) by $|\vec{p}_1+\vec{p}_2|$. Recalling that 
$\sigma\sim\sqrt{s}$ and $v_{\mbox{\scriptsize rel}}=1-\cos\theta$, one finds
\beq
\alpha=\frac{\int d^3p_1\,d^3p_2\,f(E_1)f(E_2)\,\sqrt{s}\,(1-\cos\theta)\,
|\vec{p}_1+\vec{p}_2|}{
\int d^3p_1\,d^3p_2\,f(E_1)f(E_2)\,\sqrt{s}\,(1-\cos\theta)\,(E_1+E_2)}\,.
\eeq
To get an estimate, we use the relation
\beq
|\vec{p}_1+\vec{p}_2|=\sqrt{E_1^2+E_2^2+2E_1E_2\cos\theta}
\geq(E_1+E_2)\sqrt{(1+\cos\theta)/2}\,,
\eeq
where $\theta$ is the angle between $\vec{p}_1$ and $\vec{p}_2$ and the 
equality is realized for $E_1=E_2$. Since $\sqrt{s}\sim\sqrt{1-\cos\theta}$ 
and since the angular integration is performed with the measure $d(\cos
\theta)$, the following bound can be derived:
\beq
\alpha>\frac{\int_{-1}^{1}d(\cos\theta)\,\sqrt{(1+\cos\theta)/2}\,\,
(1-\cos\theta)^{3/2}}
{\int_{-1}^{1}d\cos\theta\,(1-\cos\theta)^{3/2}}=\frac{5\pi}{32}\,.
\eeq
Thus, we find $0.5\lsim\alpha<1$, and the maximally extended $\rho^2$ 
period produces a dark energy contribution 
\beq
\Omega_d=\frac{90C}{g_*\pi^2}\,\cdot\,\frac{\alpha}{3}\ln\left(M_4\lc/12^{\,
3/2}\right)
\,,
\eeq
where $C$ is defined in Eq.~(\ref{cdef}). Even for the conservative number 
$\alpha\simeq 0.5$, this clearly dominates over the contribution from the 
$\rho$ period and corresponds, in the most optimistic scenario with $\lc 
\simeq 1\,\mbox{mm}=1/(0.2\,\mbox{meV})$, to $\Omega_{d,N}\simeq 0.05$. Such 
a value is marginally consistent with present nucleosynthesis bounds and 
would probably be visible in future CMBR analyses. 

As we discuss in the next section, the role of the black hole in late 
cosmology has a nice interpretation in the AdS/CFT correspondence. 
However, the $\rho^2$-period depends crucially on which definition of 
RSII cosmology is chosen -- that inherited from the AdS/CFT correspondence, 
or the above naive `brane in AdS$_5$' definition.

\section{The AdS/CFT definition}

Many of the confusions concerning the early cosmology of RSII stem from
the fact that there are various definitions of the theory, which
accord with each other in the IR, but which can differ in their high-energy
and high-temperature behaviour.  The first definition is that employed
by RSII and is the one used in the previous sections.  Specifically, this 
involves a brane whose motion continues to be well-described by
five-dimensional gravity at length scales below the AdS length $\lc$, the
5d gravity description only breaking down at the fundamental 5d
Planck length $1/\mfi$. 

The second definition, which we now discuss, arises from the
AdS/CFT correspondence \cite{mald,witten,gkp}.  Although more
complicated in detail, 
this definition has the significant advantage that it is embedded
in a richer context, allowing in principle a self-consistent discussion
of high-energy (and high-temperature) effects via the
connection to string theory. 

Let us recall some basic aspects of the AdS/CFT
correspondence in the most studied case. IIB string theory on
AdS$_5\times$S$^5$ is conjectured to be equivalent to an ${\cal N}=4$
supersymmetric Yang-Mills conformal field theory on the boundary of
AdS$_5$.  Defining the 't~Hooft coupling of the boundary $SU(N)$ SYM theory
$\lambda = g^2_{YM} N$, the relation between the parameters of the
theories is
\beq
g_{YM}^2 = 4\pi g_s\quad,
\qquad \left({\lc\over \lc_s}\right)^4 =
\left({R_{S^5}\over \lc_s}\right)^4 = \lambda\,\,,
\label{relations}
\eeq
where $g_s$ and $\lc_s$ are the string coupling and length, and
$R_{S^5}$ is the radius of the S$^5$. 
That the boundary of the full 10-dimensional space is just
4-dimensional is easily seen from inspecting the AdS$_5\times$S$^5$ metric
\beq
ds^2 = {\lc^2\over z^2}\left\{dz^2 +dx^2 + z^2 d\Omega_5^2\right\}
\label{tenDmetric}
\eeq
in the limit $z\to 0$.  Note that the radius of the S$^5$ part of the
background, as measured by an observer located at $z=\lc$, is $\lc$.
This feature is quite generic, applying to the broad class of backgrounds
of the form AdS$_5\times$M$_5$, since it derives from the ballancing of 
curvature contributions from the AdS$_5$ and M$_5$ in the bulk
equations of motion.
 
The precise fashion in which these two theories correspond
is most easily described in terms of the generating functional of
correlation functions for the boundary theory
\beq
Z[\psi(x)]=\left\langle
\exp\int dx \psi(x){\cal O}(x)\right\rangle \,
\eeq
where ${\cal O}(x)$ are a complete set of operators of the boundary
theory, and $\psi$ are, for the moment, just the associated sources.
The proposal of Refs.\cite{gkp,witten} is to identify $Z[\psi(x)]$
with the functional integral of the AdS$_5\times$S$^5$ theory, with
specified boundary behaviour for the fields: 
\beq
Z[\psi(x)] = \int [d\Psi(z,x)] \exp\left( -S_{\mbox{\scriptsize IIB}}[\Psi(
z,x)]\right)\biggr|_{\Psi(z\to 0,x) \to \psi(x)} \,\,\,.
\label{functional}
\eeq
In the limit of large $N$ and large 't~Hooft coupling $\lambda$, the bulk
IIB string theory goes over to its (super)gravity limit in the tree
approximation.  In fact, a similar correspondence with a 4d boundary CFT
is believed to apply much more generally for any gravity theory formulated
on AdS$_5$.  In particular we will, for simplicity, consider the 
non-supersymmetric case in the following.

It is important that, because of the short-distance singularities of the
correlators, the above expressions must be regulated so that they are 
well-defined.  A natural way of doing this is to impose the boundary 
conditions at a finite cutoff $z=z_c$ rather than at $z=0$.  
By the Weyl rescaling symmetry of the coordinates $(z,x)$, changes in
the $z$ coordinate of AdS$_5$ can be reinterpreted as changes in 
the energy scale of the CFT process under consideration, with
$z\to 0$ corresponding to the UV limit of the boundary conformal theory.
Thus the cutoff at $z=z_c$ implements a UV regulation of the
boundary theory.  In order to get a well-defined limit as the cutoff
is removed, $z_c\to 0$, the boundary theory must be supplemented by
counterterms depending on the boundary values of the bulk fields. Moreover,
the cutoff at $z=z_c$ also allows the normalizability of the bulk AdS
graviton zero mode, the existence of such a mode implying that the
regulated boundary theory now also couples to dynamical gravity. 
In the limit of $z_c\to 0$, the boundary theory
becomes a pure CFT decoupled from gravity, the effective Planck mass
of the boundary theory $\mfo$ going to infinity
as the graviton zero mode becomes non-normalizable.
  
In the specific case we are interested in, this cutoff is implemented by 
a physical `Planck brane' located at $z_c$. This Planck
brane provides a physical cutoff for the bulk theory and
the theory on the Planck brane is a 4-dimensional CFT coupled to
dynamical 4-dimensional gravity arising from the graviton zero mode.
The CFT degrees of freedom are just the new expression of the bulk
Kaluza-Klein modes of the graviton in this 4d `holographic' approach.

Following Refs.\cite{hs,gub,gkr},
it is useful to be more explicit about the result of this procedure in
the simple case where the only bulk field is the 5-dimensional metric 
$g_5(x,z)$, with boundary value $g_4(x)$ at the Planck (regulator) brane 
at $z_c$.\footnote{In addition, we suppose that there are fields $\phi$
localized to the Planck brane which represent the standard model degrees
of freedom. In general, string theory imposes strong consistency requirements
relating the spectrum of brane-localized fields to the bulk field content,
so our assumption of $g_5$ as the sole bulk field is not really justified.  
However it is sufficient to illustrate the point at hand.}
The boundary operator for which $g_4$ is the source is just the
energy-momentum tensor $T^{\mu\nu}_{\rm CFT}$ of the CFT, so
the integration over all bulk metrics will lead to the generating
functional $Z[g_4]=\langle \exp(-\int g_{4,\mu\nu}T^{\mu\nu}_{\rm CFT} ) 
\rangle_{\rm CFT}$ for correlation functions of $T_{\rm CFT}$.  In addition, 
coun\-ter\-terms in the boundary metric $g_{4,\mu\nu}$ must be added. 
Together this leads to a boundary theory described by the effective action 
\beq
S_{\rm 4d}[g_4,\phi] = \int d^4x \sqrt{g_4} \,\biggl\{\,{\cal L(\phi)
}-{1\over 2}\mfo^2{\cal R} - b_4 {\cal R}_2 + \ldots\biggr\} \ .
\label{bdryS}
\eeq
Here ${\cal R}_2 = -{\cal R}^{\mu\nu} {\cal R}_{\mu\nu}/8 + {\cal R}^2/24$
is the leading higher-derivative pure metric piece of the counterterm action,
and $b_4 = c f(z_c)$ where $f(z_c)$ is a function that
has a $\log(z_c)$ singularity as $z_c\to 0$, and
$c=2\pi^2(\mfi\lc)^3$ is the central charge of the boundary CFT.

At long distances this effective action correctly reproduces the
gravitational potential between two masses located on the Planck brane
including leading corrections:
\beq
V(r) = {m_1 m_2 \over \mfo^2} \left(\frac{1}{r}+a\frac{\lc^2}{r^3}+\ldots
\right)\ ,
\eeq
for $r\gg \lc$, and $a$ a numerical coefficient.  In
the original bulk AdS$_5$ picture, the $1/r^3$ correction
is due to the exchange of the continuous spectrum of Kaluza-Klein
modes, while in the boundary CFT + 4-dimensional gravity description,
the correction is due to the two-point function of the CFT energy-momentum
tensor with two external 4d graviton propagators
$(1/p^2)\langle T(p)T(-p)\rangle(1/p^2) \sim \log p^2$.
Recall that the two-point function of $T_{\rm CFT}$ satisfies
\beq
\langle T(p)T(-p)\rangle \sim c \  p^4 \log p^2
\eeq
with $c$ the central charge.

Note that an inspection of the calculation of Ref.~\cite{gkp} shows that
for fixed 4d Planck mass, $\mfo$, the central charge $c/2\pi^2=(\mfi\lc)^3=
(\mfo\lc)^2$ of the boundary CFT depends only on the geometry of the AdS$_5$
(through the parameter $\lc$). In particular, for equal values of $\lc$, the 
two theories defined by taking the geometry to be respectively either 
AdS$_5$ or AdS$_5 \times$S$^5$ have the {\em same} 4d CFT + 4d gravity 
description in the IR.\footnote
{
In fact, because of the isometries of S$^5$, the IR CFT of the 
AdS$_5\times$S$^5$ has additional conserved currents. However, the 
central charge and thus the late cosmology of the two models is the same
(cf.~Eq.~(\ref{rct})).
}
Since the theories most certainly differ for length 
scales $r<\lc$ (the static gravitational potential behaving as $V(r)\sim 
1/r^2$ or $\sim 1/r^7$ in the two cases), the 4d CFT + gravity description 
must break down at $r\simeq\lc$ due to the existence of new strong couplings 
possibly involving new degrees of freedom.

As pointed out by Gubser~\cite{gub}, the CFT perspective is particularly 
useful for a physical understanding of the dark radiation term in the 
Friedmann equation. The AdS-Schwarzschild geometry implies, by continuation 
to Euclidean space, a black hole temperature $T_{\rm BH}=r_h/\pi\lc^2$. This
corresponds to a local temperature $T(r)=T_{\rm BH}/\sqrt{f(r)}$ in the 
bulk~\cite{hp}. For $R\gg r_h$, the temperature near the brane is $T(R) 
\simeq r_h/(\pi\lc R)$. Thus, the brane observer interprets the CFT, 
represented by the bulk degrees of freedom, as being heated to this 
temperature. At weak coupling, the energy density of the heated CFT is 
given by $\rho=2\pi^2cT^4$. Taking into account the famous factor $3/4$
(see, e.g.,~\cite{itz}), which is due to strong coupling effects in the CFT,
one finds 
\beq
\rho=\frac{3\pi^2}{2}c\,[T(R)]^4\,,\label{rct}
\eeq
which is precisely the dark radiation term of Eq.~(\ref{hu}). Note that, 
phenomenologically, $T(R)$ has to be much smaller than the standard model 
temperature on the brane. Otherwise, the dark radiation completely
dominates the total energy density due to the large number of degrees of 
freedom of the CFT. 

Near the horizon, $f(r)$ goes to zero and the local bulk temperature 
diverges. Nevertheless, the brane observer in a cosmological setting sees 
an effective bulk temperature $T=r_h/(\pi\lc R)$, since the red-shift 
factor from the brane motion in the black-hole rest frame compensates the 
singular behaviour of $1/\sqrt{f(r)}$. We have, at present, no deeper 
understanding of this intriguing fact. Note, however, that the naive 
formula $T(r)=T_{\rm BH}/\sqrt{f(r)}$, which underlies the above discussion, 
should in itself be questioned near the horizon, since it is in principle
affected  by the ambiguities of the definition of the black-hole vacuum state 
(cf.~\cite{bd}). 

In the cosmological context of the earlier sections, an immediate question
is if the $\rho^2$ correction in the Friedmann equation~(\ref{hu}) is 
similarly 
reproduced by the AdS/CFT description Eq.(\ref{bdryS}). At first glance the 
answer appears to be no, as the $\phi$-matter energy-momentum tensor
derived from ${\cal L}(\phi)$ couples linearly with $g_{4,\mu\nu}$. However, 
the existence of the 
higher-derivative terms ${\cal R}_2 = -{\cal R}^{\mu\nu}{\cal R}_{\mu\nu}/8
+{\cal R}^2/24$ in the 4d action Eq.(\ref{bdryS}) leads to an effective
$\rho^2$ correction.\footnote
{
After completion of this paper we were kindly informed of the recent work of 
Ref.~\cite{si}, where the correct coefficient of the $\rho^2$ term in the 
pure AdS$_5$ case has been verified. 
}
(Recall that these higher-derivative terms are
predicted by the conformal anomaly of the boundary CFT and occur with
coefficient $b_4$ enhanced by $c$.)  To see this, consider the Friedmann 
equations in the limit that $\rho\ll \mi^4$; in this case the Hubble length 
$H^{-1}\gg \lc$.  Then in the leading approximation $H^2\sim\rho/\mfo^2$, 
and substituting this back in to the higher-derivative ${\cal R}^2$ terms 
gives
\beq
b_4\biggl(-{\cal R}^{\mu\nu}{\cal R}_{\mu\nu}/8 +{\cal R}^2/24\biggr)
\sim{c H^4} \sim {\lc^2\rho^2\over\mfo^2} = {\rho^2\over\mi^4} \ ,
\eeq
the required $\rho^2$ term.  The coefficient of this effective
$\rho^2$ correction can vary, though, depending on the definition of the
UV theory.  For example, the boundary action Eq.(\ref{bdryS}) can contain
an additional higher-dimensional counterterm that is not forbidden by any 
symmetry:
\beq
{1\over \mfo^4}\, T_{\mu\nu}^{\rm CFT} T^{\mu\nu}_\phi \ .
\eeq
Such an operator directly linking the CFT with the brane-localized
matter has recently been invoked \cite{ahpr} to explain the $1/r^7$
corrections to the gravitational potential found in the Lykken-Randall 
`probe-brane' scenario \cite{lr} (see also~\cite{gk}). Via the use of the 
$T_{\rm CFT}$ two-point 
function, this operator in turn implies the additional interaction
\beq
\Delta S_{\rm 4d} \sim {1\over \mfo^8}\int d^4x d^4y T_\phi(x) \langle 
T_{\rm CFT}(x)T_{\rm CFT}(y)\rangle T_\phi(y)\sim {c\over \mfo^8} \int d^4x 
d^4y{T_\phi(x) T_\phi(y) \over |x-y|^8} 
\eeq
which, for homogeneous and isotropic radiation $T_\phi={\rm diag}(\rho,-
\rho/3,-\rho/3,-\rho/3)$, leads to a term (taking the cutoff for the Planck 
brane theory to be $\mfo$)
\beq
{c\over \mfo^4}\rho^2 \sim {\rho^2\over \mi^4}
\eeq
on the rhs of the Friedmann equation. 

Thus it is possible to account for the
$\rho^2$ term in the modified Friedmann equation, Eq.(\ref{hu}), from a CFT
description, although the exact coefficient of this term depends on the UV
definition of the theory.  In the case that the UV ($r<\lc$) theory is 
just the brane in AdS$_5$, then, by definition, the CFT version must exactly 
reproduce the term in Eq.(\ref{hu}).  Alternatively if the theory is the
string-motivated AdS$_5\times$S$^5$, then the $\rho^2$ term can receive
further corrections.

Despite the existence of the $\rho^2$ correction to the Friedmann equation,
we wish to emphasize that the unusual cosmological 
`$\rho^2$-behavior' where this term dominates (occurring for $\rho> \mi^4$)
is not physically accessible in the string motivated
AdS$_5 \times$S$^5$ definition of the theory.  The reason for this is simply
that for the string coupling $g_s <1$ (which can always be arranged by
suitable strong to weak coupling duality transforms) the string scale 
satisfies $L_s^{-1}\sim g_s^{1/4}(\mfo/\lc^3)^{1/4}\sim g_s^{1/4}(\mi^3/
\mfo)^{1/2} \ll \mi$.  Thus one encounters the full string theory well 
before one reaches the scale at which $\rho^2$ behaviour begins.

\section{Summary and Comments}

In this paper we have considered some of the cosmological
issues raised by taking the Randall-Sundrum~II framework to
describe our world.  The Randall-Sundrum~II proposal assumes that
the SM degrees of freedom are localized on the `Planck brane' 
and are therefore described by a conventional (3+1)-d quantum field
theory up to high scales; thus the successes of, for example,
supersymmetric gauge coupling unification are not necessarily spoiled.
However, RSII does not itself provide a solution to the hierarchy
problem -- one must still assume supersymmetry or technicolour, or
some other as yet undiscovered mechanism, to solve this problem for
the Planck-brane QFT; on the other hand, gravity is dramatically
modified leading to new effects, especially in early universe cosmology,
and associated constraints or signals.  RSII provides an interesting
laboratory for exploring modifications to gravity and early universe
cosmology.

There are two immediate new effects in RSII cosmology.  The first
is that the motion of our Planck brane in the AdS bulk leads to 
a correction to the usual Friedmann equations for the scale
factor -- the well known $\rho^2$-term.  The second is
the possibility of radiation into the AdS bulk, and the
formation of a black hole in the AdS spacetime.  Both of these issues
have been previously studied. However, we argue that the amount of `dark 
radiation' that could be produced during the linear-in-$\rho$ and $\rho^2$ 
epochs in the early universe was overestimated in previous treatments. 
This leads to the false conclusion that the Planck
brane temperature could never be high enough to reach the $\rho^2$
region. We give a detailed account of this radiation and further
show that the formation of the bulk AdS-Schwarzschild background
is a necessary consequence. The numbers that we find for the dark radiation 
are in a region that is accessible by future more precise cosmological 
measurements. In particular, we predict dark radiation at nucleosynthesis 
with $\Omega_{d,N}\simeq 0.005$ from the $\rho$ regime and $\Omega_{d,N} 
\lsim 0.05$ from the $\rho^2$ regime. The precise amount of the contribution 
from the $\rho^2$ regime depends on the maximal brane temperature in the 
early universe. 

Another important ingredient of our paper is the extended discussion
of RSII cosmology in the framework of the AdS/CFT correspondence. 
We show how, for small brane energy densities, the manifestly
4-dimensional CFT description can reproduce the predictions
of the standard definition in which the Planck
brane moves in AdS$_5$ according to the Israel junction
conditions, including the $\rho^2$ corrections to the Friedmann
equations.  However, we emphasize that there exist different 
definitions of the theory in the `UV region' $r<\lc$, and the
precise size of the $\rho^2$ corrections depends on
this UV theory.  In the general case, the corrections to the
cosmological evolution arise both from higher-derivative terms in
the curvature and higher-dimension operators connecting the 
on-brane matter directly to the CFT.  Because of the
large central charge of the CFT, the coefficient of the higher
curvature terms is greatly enhanced over its naive value, indicating
a precocious breakdown of the 4d effective theory.  
In the case of the string-motivated definition on AdS$_5\times$M$_5$,
the `$\rho^2$-region' where the $\rho^2$ term dominates the evolution
is not physically accessible.  

There are many additional physical questions that either deserve
further attention or have not been discussed at all.  These
include a further analysis of the near horizon,
$(R-r_h)/r_h\ll 1$, behavior of the `Israel' brane theory, and
the inclusion of the many more bulk fields (in addition to the graviton)
that are likely to accompany the construction of even a semi-realistic
model.  At low energies, the manifestly 4-dimensional
CFT+gravity description is likely to be useful to analyse such
questions as the production and evolution of the cosmic
microwave background fluctuations.  Concerning the UV definition
of the RSII theory, an important general point
about the string-motivated AdS$_5\times$M$_5$ definition is that it
encodes a consistent description of the theory in the
high-energy regime, with different physical behaviour from the
standard `Israel' definition.

\section*{Acknowledgements}
We would like to thank Tony Gherghetta, Yaron Oz and Riccardo
Rattazzi for helpful discussions. We are also grateful to the authors 
of the later paper Ref.~\cite{lsr} (see also~\cite{ls}) for pointing out 
two numerical errors in the first version of this paper. Special thanks go 
to Lorenzo Sorbo, who helped to compare the respective calculations.

\section*{Appendix}
\setcounter{equation}{0}\renewcommand{\theequation}{A.\arabic{equation}}

In this appendix, we briefly describe the calculation of the cross sections 
for the production of bulk gravitons by scalars, fermions and vector 
particles on the brane. The effective 4-dimensional Lagrangian for the 
graviton Kaluza-Klein modes and their coupling to matter has been derived,
e.g., in Ref.~\cite{ced}. To properly normalize the bulk excitations, one 
introduces a regulator brane at $r=\lc\exp(-L/\lc)$ and takes the limit
$L\to\infty$ at the end. In the transverse traceless gauge, where the 
graviton fields $H^{\mu\nu}_{(n)}$ satisfy $H^{\mu\nu}_{(n)\,,\,\mu}
=0$ and $g_{\mu\nu}H^{\mu\nu}_{(n)}=0$, the Lagrangian reads\footnote{
Note that we have changed the normalization of $H_{\mu\nu}$ by a factor 
of $\sqrt{2}$ with respect to Ref.~\cite{ced} to give the graviton fields 
a canonical kinetic term.}
\beq
{\cal L}_H=-\frac{1}{2}\sum_n\left(H^{\mu\nu}_{(n)\,,\,\rho}H_{\mu\nu}^{(n)
\,,\,\rho}+m_n^2H^{\mu\nu}_{(n)}H_{\mu\nu}^{(n)}\right)+T^{\mu\nu}\sum_n
\lambda_nH_{\mu\nu}^{(n)}\,,\label{lag}
\eeq
where, in the limit $m_n\gg 1/\lc$,
\beq 
m_n=\frac{1}{\lc}\,n\pi e^{-L/\lc}\qquad\mbox{and}\qquad\lambda_n=
\frac{1}{M_4}\,e^{-L/2\lc}\,.
\eeq
For each $n$, the operator $H_{\mu\nu}^{(n)}$ represents simply a massive
spin-2 field with 5 physical degrees of freedom (see, e.g.,~\cite{velt}). 
Thus, introducing the 5 polarization tensors $\epsilon_{\mu\nu}^{(i)}(q)$ 
(with $\epsilon_{\mu\nu}^{(i)}(q)\epsilon^{\mu\nu}_{(j)}(q)=\delta^i_j$), 
the final state spin sum in the production of a graviton with momentum $q$ 
can be taken multiplying the squared amplitude with 
\beq
\sum_i\epsilon_{\mu\nu}^{(i)}(q)\epsilon_{\alpha\beta}^{(i)}(q)=
g_{\mu\alpha}g_{\nu\beta}-\frac{1}{3}g_{\mu\nu}g_{\alpha\beta}\,\,+\,\,
(\,\mbox{terms}\,\,\sim q_\mu\,,\,q_\nu\,,\,q_\alpha\,\,\mbox{or}\,\,
q_\beta\,)
\eeq
and summing over $\mu\nu$ and $\alpha\beta$. Terms $\sim q$ are irrelevant 
since the energy-momentum tensor $T^{\mu\nu}$ in Eq.~(\ref{lag}) is 
conserved. 

Since we will mainly be interested in cms-energies in the TeV region, we 
disregard the masses of brane degrees of freedom in the following 
calculation. In the case of one real scalar field, the energy-momentum 
tensor in Eq.~(\ref{lag}) is given by 
\beq
T_{\mu\nu}=\partial_\mu\phi\partial_\nu\phi-\frac{1}{2}g_{\mu\nu}(\partial
\phi)^2\,.
\eeq
Now it is straightforward to derive the Feynman rule for the $\phi\phi H$ 
vertex and to calculate the total $\phi\phi$ annihilation cross section 
after the sum over $n$ is replaced by an integral over the final state mass 
$m=\sqrt{s}$ according to 
\beq
\sum_n\,\,\longrightarrow\,\, \frac{\lc}{\pi}\,e^{L/\lc}\,\int d\sqrt{s}\,.
\eeq
Similarly, the annihilation cross sections for two vector particles (e.g., 
two photons) and for a fermion-antifermion pair can be obtained by 
substituting the energy-momentum tensors
\beq
T_{\mu\nu}=-F_{\mu\rho}F_{\nu}^{\rho}+\frac{1}{4}g_{\mu\nu}F^2\,,
\eeq
with $F_{\mu\nu}=\partial_\mu A_\nu-\partial_\nu A_\mu$, and
\beq
T_{\mu\nu}=\frac{i}{4}\bar{\psi}\left(\gamma_\mu\stackrel{\leftrightarrow}
{\partial}_\nu+\gamma_\nu\stackrel{\leftrightarrow}{\partial}_\mu\right)\psi
\eeq
into Eq.~(\ref{lag}). The resulting cross sections are given in 
Eqs.~(\ref{cs}) and (\ref{cfa}) of the main part of the paper. 

Note that, since we are only interested in the limit $\sqrt{s}\gg 1/\lc$, 
the AdS nature of the bulk has no physical significance. Therefore, our 
result can be compared with the graviton production cross section on a
brane in a flat, 5-dimensional bulk (the ADD scenario with one extra 
dimension~\cite{add}) if the 5-dimensional gravitational constants are 
identified. Indeed, one can check explicitly that the one-graviton exchange 
amplitude calculated in the framework of this appendix (where $i/(-q^2-m_n^2
)\,\sum_j\epsilon_{\mu\nu}^{(j)}(q)\epsilon_{\alpha\beta}^{(j)}(q)$\,) is 
the graviton propagator) is identical to the result of Ref.~\cite{grw} 
(see Eqs.~(69)--(71) with $\delta=1$).

\end{document}